\documentclass[twocolumn, pra]{revtex4-1}

\usepackage{natbib}
\usepackage{graphicx}
\usepackage{dcolumn} 
\usepackage{bm}
\usepackage{hyperref}
\usepackage{mathtools}
\usepackage{amssymb}
\usepackage{enumitem}
\usepackage[mathlines]{lineno}
\usepackage[titletoc, title]{appendix}

\begin{document}

\title{Vortices with massive cores in a binary mixture of Bose-Einstein condensates }

\author{Andrea Richaud}
\email{Correspondence to: andrea.richaud@polito.it}
\affiliation{
Dipartimento di Scienza Applicata e Tecnologia and u.d.r. CNISM, Politecnico di Torino, 
Corso Duca degli Abruzzi 24, I--10129 Torino, Italy}

\author{Vittorio Penna}
\affiliation{
Dipartimento di Scienza Applicata e Tecnologia and u.d.r. CNISM, Politecnico di Torino, 
Corso Duca degli Abruzzi 24, I--10129 Torino, Italy}

\author{Ricardo Mayol}
\affiliation{Departament de F\'{\i}sica Qu\`{a}ntica i Astrof\'{\i}sica and Institut de Ci\`{e}ncies del Cosmos,
Facultat de F\'{\i}sica, Universitat de Barcelona, E--08028 Barcelona, Spain}

\author{Montserrat Guilleumas}
\affiliation{Departament de F\'{\i}sica Qu\`{a}ntica i Astrof\'{\i}sica and Institut de Ci\`{e}ncies del Cosmos,
Facultat de F\'{\i}sica, Universitat de Barcelona, E--08028 Barcelona, Spain}

\date{\today}

\begin{abstract}
We analyze a notable class of states relevant to an immiscible bosonic binary mixture loaded in a rotating box-like circular trap, i.e. states where vortices in one species host the atoms of the other species, which thus play the role of massive cores. Within a fully-analytical framework, we calculate the equilibrium distance distinguishing the motion of precession of two corotating massive vortices, the angular momentum of each component, the vortices healing length and the characteristic size of the cores. We then compare these previsions with the measures extracted from the numerical solutions of the associated coupled Gross-Pitaevskii equations. Interestingly, making use of a suitable change of reference frame, we show that vortices drag the massive cores which they host thus conveying them their same motion of precession, but that there is no evidence of tangential entrainment between the two fluids, since the cores keep their orientation constant while orbiting.

\end{abstract}

\maketitle

\section{Introduction}
\label{sec:Introduction}
Vortices in quantum fluids are topological excitations characterized by quantized circulation \cite{Onsager_Cimento} which are present in a number of nonlinear field theories and models \cite{Pismen}, ranging from superfluid media \cite{Feynman_Low_T,Donnelly} and quantum optics \cite{Brambilla_Penna,Carusotto} to superconductivity theories \cite{Blatter_vortices,Huebener} and Josephson-junction arrays \cite{Josephson_Vortices,Fazio_QPTs} and play a key-role in fundamental effects such as superfluid turbulence \cite{Barenghi_Donnelly}, the Berezinskii-Kosterlitz-Thouless transition \cite{Kosterlitz_Thouless}, fractional statistics \cite{Chiao_Hansen}, and in the development of a fully quantized field theory for topologically complex excitations \cite{Fetter_1967,Rasetti_Regge,Penna_Rasetti_Spera}. Among the plethora of different physical systems where vortices can be experimentally investigated, ultracold quantum gases provide a particularly controllable and versatile platform \cite{Fetter_review,Zoller_Vortices_Solitons} for the study and the observation of the rich phenomenology associated to their formation \cite{Butts,Dalfovo_formation}, dynamics \cite{Pismen_dynamics,Stringari_lattice_dynamics}, and interactions \cite{Castin_Dum}. Vortices in Bose-Einstein condensates (BECs) were first obtained by means of a phase-imprinting method involving two hyperfine spin states of $^\text{87}$Rb \cite{Matthews_first_vortex} but, at present, can be produced also by stirring the BEC above a certain critical velocity \cite{Overcritical_rotation,Sinha,Rotating_Dalibard}, dragging barriers through the BEC itself \cite{Onofrio} or interfering multiple condensate fragments \cite{Scherer}. 

Solitons are a kind of localized excitations which, because of the competition between dispersion and nonlinearity, propagate keeping their shape unaltered, even if a two-soliton collision occurs \cite{Drazin}. Soon after the achievement of Bose-Einstein condensation, different types of solitons have been described and observed \cite{Zoller_Vortices_Solitons,Burger,Frantzeskakis}. To our purposes, of particular importance are those systems where a bosonic binary mixture features dark-bright soliton configurations \cite{Becker,Hamner,Middelkamp}. These structures, first predicted in Ref. \cite{Busch_Anglin},  are frequently termed as ``symbiotic solitons" \cite{Symbiotic_solitons} because the bright component, being endowed with repulsive intraspecies interaction, could not exist if the dark component did not play the role of an effective confining potential. 

The same symbiotic relationship was shown to constitute the mechanism underlying the robustness of vortex-bright soliton complexes. First observed in 2000 by the JILA group \cite{JILA}, they represent the topological extension of the dark-bright soliton configuration to the case where a component hosts one or more vortices \cite{Kevrekidis_PRL}. The aforementioned study paved the way to a series of further investigations which highlighted, among various aspects, the spontaneous generation of vortex-bright soliton structures \cite{MukherjeeQuench}, the possibility, for the effective potential well corresponding to the vortex core, to support not only bound states \cite{Gallemi_Stringari}, but also multi-ring excited radial state complexes \cite{Charalampidis}, and a rich dynamical scenario for the bright-solitary component \cite{Vortices_double_well}. 

Within an analytical framework and by means of extensive numerical simulations, our work aims at analyzing the static and the dynamical properties of vortex-bright soliton complexes, i.e. how the presence of massive solitons within the cores of two corotating vortices affect the equilibrium distance characterizing their motion of precession around the trap center, the role of the interspecies repulsion as an antagonist to the centrifugal force acting on the solitons, the functional dependence of the angular momenta carried by two species, of the vortex healing length, and of the characteristic radius of the massive cores on the mass of the latter. 

If one considers repulsive intraspecies ($g_a$, $g_b$) and interspecies ($g_{ab}$) interactions such that the immiscibility condition $g_{ab}>\sqrt{g_a g_b}$ is fulfilled, the dynamical picture of the mixture [in which the order-parameter fields of the two species obey two coupled Gross-Pitaevskii equations (GPEs)], indeed reduces to the much simpler equations of two point-like vortices with nonzero-mass cores. Noticeably, the latter are found to exhibit an evident Lorentz-like form since, in the presence of vortex cores occupied by a second species, the vortex-motion equations are equivalent to those of a pair of massive charges acted by a transverse magnetic field. With negligible fractions of the minority component, one recovers the Helmholtz-Kirchhoff equations for planar point-like vortices \cite{Saffman}.

The outline of the manuscript is the following: in Sec. \ref{sec:Analytical_model}, we present an analytical model for the dynamics of massive vortices in a confined system which incorporates the effect of the virtual vortices resulting from the boundary condition of vanishing normal velocity. In particular, we derive a formula giving the equilibrium distance distinguishing the motion of precession of two corotating massive vortices. Sec. \ref{sec:Numerical_model} is devoted to the presentation of the two coupled stationary GPEs which provide a good description of the bosonic binary mixture in the mean-field approximation. In Sec. \ref{sec:Vortices_BECs}, we show how the presence of massive cores (i.e. species-\textit{b} atoms trapped within species-\textit{a} vortices) affects the equilibrium distance of the pair of corotating vortices. We also show that the interspecies repulsion tends to counterbalance the centrifugal force acting on species-\textit{b} atoms. In Sec. \ref{sec:Angular_momentum}, we address the angular momenta of the two components and provide analytical formulas that well capture their functional dependence on the number of species-\textit{b} atoms (which, in turn, is directly proportional to the mass of the cores). By means of a suitable change of reference frame, we show that the cores, although following the same motion of precession of the vortices, keep their orientation constant while orbiting. This circumstance witnesses the fact that there is no tangential entrainment between the two fluids. In Sec. \ref{sec:Healing_diameter}, we present an heuristic but effective system of equations that well reproduces the functional dependence of the vortex healing lengths and of the cores' characteristic radius on the number of species-\textit{b} atoms. Eventually, Sec. \ref{sec:Conclusions} is devoted to concluding remarks.

\section{point-like vortices in a circular box}
\label{sec:Analytical_model}
In this section, we review some results concerning the dynamics of point-like vortices and we introduce a model for the dynamics of vortices whose cores host point-like masses (hence the name \textit{massive} vortices, as opposed to the traditional \textit{massless} vortices).
\subsection{Massless vortices}
\label{subsec:Massless_vortices}
The Hamiltonian of $N$ point-like massless vortices in an ideal unbounded fluid is given by \cite{Saffman} 
\begin{equation}
\label{eq:Hami_unbounded}
      H_\infty=(z_1,\,\dots,\,z_N)=-\frac{\rho_*}{4\pi}\sum_{i=1}^N \sum_{j\neq i} k_i k_j \ln \frac{|z_i-z_j|}{\lambda}, 
\end{equation}
where $\rho_*$ is the fluid planar density, $z_j=x_j+iy_j\,\in\,\mathbb{C}$ is the position of the $j$th vortex in the ambient plane and $k_j=n_j h/m_f$ is its strength ($n_j\,\in\,\mathbb{Z}$ is the vortex quantization and $m_f$ is the mass of the fluid particles). In the following, we will specialize our discussion to the case of $N=2$ vortices.

When one considers bounded systems, Hamiltonian (\ref{eq:Hami_unbounded}) modifies due to the presence of the confining potential. In the case of a box-like potential (this type of confinement is within the reach of current experimental trapping techniques, see, e.g., Refs. \cite{DMD1,DMD2,Minardi}), the presence of a boundary confining the fluid is accounted for by means of the virtual charge method, i.e. by introducing a suitable configuration of virtual vortices. With this premise in mind, the Hamiltonian of $N=2$ point-like massless vortices in an ideal fluid confined in a circular box of radius $R$ reads \cite{Penna_box}
$$
  H=\frac{\rho_*}{4\pi}\left\{k_1k_2 \log \frac{|R^2-z_1\bar{z}_2|^2}{|R(z_1-z_2)|^2} \right.
$$
\begin{equation}
\label{eq:Hami_bounded}
   \left. +k_1^2\log\left(1-\frac{|z_1|^2}{R^2}\right)+k_2^2\log\left(1-\frac{|z_2|^2}{R^2}\right) \right\}.  
\end{equation}
In this framework, the coordinates of each vortex constitute a pair of canonically conjugate variables, and motion equations can be obtained by means of the Poisson Brackets
$$
  \{F,\,G\}= \frac{1}{\rho_* k_j }\sum_{j=1}^2   \left[\frac{\partial F}{\partial x_j} \frac{\partial G}{\partial y_j}- \frac{\partial G}{\partial x_j}\frac{\partial F}{\partial y_j}\right ]
$$
involving, in turn, the canonical brackets $\{x_i,y_j\}=\delta_{i,j}/(\rho_* k_j)$ (see for example \cite{Penna_PRB_59}). 

\subsection{Massive vortices}
\label{subsec:Massive_vortices}
If one wants to introduce into the model the fact that the vortex cores host point masses, it is convenient to move to the Lagrangian formalism, where the presence of massive cores can be taken into account as follows
\begin{equation}
\label{eq:Lagrangian}
 L= \sum_{j=1}^2 \left[ \frac{m_j}{2}(\dot{x}_j^2 + \dot{y}_j^2 ) + \frac{k_j\rho_*}{2} (y_j\dot{x}_j - x_j \dot{y}_j )\right] - H,
\end{equation}
where $m_j$ represents the point-like mass hosted by the $j$th vortex core and where $\dot{q}_j:=\mathrm{d}q_j /\mathrm{d}t$ (with $q =x,y$). Note that, Lagrangian (\ref{eq:Lagrangian}) is formally equivalent to that describing charged particles in a planar domain subject to a transverse magnetic field, where $k_j$s and $\rho_*$ play the role of charges and of magnetic field, respectively. As is well known, the dynamics of the vortex cores is generated by the Euler-Lagrange equations which, in the special but interesting case of two equal vortices $k_1=k_2=k$, whose cores host two equal masses $m_1=m_2=m$, takes the form
$$
  m\ddot{\vec{r}}_j = k \rho_* \vec{u}_3\wedge \dot{\vec{r}}_j+\rho_*\frac{k^2}{2\pi}\left[\frac{\vec{r}_j-\vec{r}_i}{|\vec{r}_j-\vec{r}_i|^2}+\frac{\vec{r}_j}{R^2-r_j^2}+\right.
$$
$$
  \left. \frac{R^2\vec{r}_i-r_i^2\vec{r}_j}{R^4-2R^2\vec{r}_i\vec{r}_j+r_i^2r_j^2}\right],
$$
for $i,\,j \in \{1,2\}$ and $i \neq j$ [$\vec{u}_3$ is the unit vector perpendicular to the plane $(x,\,y)$].

The resulting system of 4 differential equations admits a notable solution, 
$$
  x_1(t)=\frac{d}{2}\cos(\Omega t), \qquad y_1(t)=\frac{d}{2}\sin(\Omega t)
$$
$$
  x_2(t)=\frac{d}{2}\cos(\Omega t+\pi), \qquad y_2(t)=\frac{d}{2}\sin(\Omega t +\pi ),
$$
provided that the two vortices are placed symmetrically with respect to the box-trap center and that their distance $d$ and the angular frequency $\Omega$ marking their motion of precession fulfill the following equation:
$$
  \frac{\pi  d^6 \Omega  (k \rho_* -m \Omega )+3 d^4 k^2 \rho_* }{d-2 R} =
$$
\begin{equation}
  -\frac{16 \pi  d^2 R^4 \Omega  (m \Omega -k \rho_* )+16 k^2 \rho_*  R^4}{d-2R}.
\label{eq:Equi_distance_implicit}
\end{equation}
As expected, Eq. (\ref{eq:Equi_distance_implicit}) shows a pathology when $d\to 2R$, meaning that the vortex pair is approaching the circular-box boundary. Moreover, in the limit of infinite box radius ($R\to + \infty$), one can retain only those terms $\propto R^4$ and the relation $d(\Omega)$ can be expressed in closed form, i.e.:
\begin{equation}
    \label{eq:d_closed_form}
    d= k\sqrt{\frac{\rho_*}{\pi}} \frac{1}{\sqrt{k\rho_*  \Omega - m  \Omega^2  }}.
\end{equation}

In Sec. \ref{sub:Equilibrium_distance}, the equilibrium distance $d$ predicted by Eq. (\ref{eq:Equi_distance_implicit}) and relevant to two equal point-like massive vortices in a circular box will be compared to the one obtained by numerically solving two coupled stationary GPEs. To conclude this Section, we would like to remark that the extension of model (\ref{eq:Lagrangian}) to the case of \textit{harmonic} confinement is far from being trivial, as the unavoidable curvature of the enveloping wavefunction produces non-negligible effective forces acting on the vortices' centers, which distort the usual vortex dynamics already in the case of zero species-$b$ atoms. It is indeed the use of a box-like potential that allows one to bypass the influence of the aforementioned non-negligible effective forces, thus allowing for a cleaner emergence of the genuine phenomenology characterizing vortex/bright-soliton complexes.

\section{The bosonic mixture}
\label{sec:Numerical_model}
We consider a bosonic mixture of $^{23}\mathrm{Na}$ and $^{39}\mathrm{K}$ \cite{Zenesini,Noi_Zenesini}. Each atomic species is characterized by an order parameter, $\varphi_a=\sqrt{N_a}\psi_a$ and $\varphi_b=\sqrt{N_b}\psi_b$ respectively. In a mean-field treatment of the problem, we assume that the system is effectively quasi-2D, as a result of a strong confinement along the $z$-direction. Because of this, it can be effectively modeled by the following two coupled stationary Gross-Pitaevskii equations 
$$
    -\frac{\hbar^2}{2m_a} \left[ \frac{\partial^2 }{\partial x^2} +  \frac{\partial^2 }{\partial y^2}   \right]
 \psi_a + \frac{g_a N_a}{\ell_z} |\psi_a|^2\psi_a  
$$
$$
 + \frac{g_{ab} N_b}{\ell_z}|\psi_b|^2\psi_a +V_{ext,a} \psi_a  =\mu_a \psi_a 
$$
$$
-\frac{\hbar^2}{2m_b}\left[ \frac{\partial^2 }{\partial x^2} +  \frac{\partial^2 }{\partial y^2} \right]  \psi_b + \frac{g_b N_b }{\ell_z} |\psi_b|^2\psi_b
$$
\begin{equation}
\label{eq:Eig_prob}
       + \frac{g_{ab}N_a}{\ell_z}|\psi_a|^2\psi_b +V_{ext,b} \psi_b  =\mu_b \psi_b 
\end{equation}
where $N_a$ ($N_b$) corresponds to the number of species-\textit{a} (species-\textit{b}) atoms, $g_c=4\pi \hbar^2 a_c /m_c $, with $c=a,\,b$ are the intraspecies interaction strengths and  $g_{ab}=2\pi \hbar^2 a_{ab} /m_{ab} $ is the interspecies coupling. Notice that $m_a$ ($m_b$) is the atomic mass of sodium (potassium), while $m_{ab}=(m_a^{-1}+m_b^{-1})^{-1}$ is their reduced mass; similarly $a_a$ and $a_b$ are the intraspecies scattering lengths, while $a_{ab}$ is the interspecies scattering length. Parameter $\ell_z$ is the effective thickness of the disk-like box trap and functions $\psi_a$ and $\psi_b$ are normalized to $1$, since $\varphi_a$ and $\varphi_b$ are, respectively, normalized to $N_a$ and $N_b$.

Vortical solutions of Eqs. (\ref{eq:Eig_prob}) are found by moving to a frame rotating with angular velocity $\Omega$ (this corresponds to adding the term $-\Omega \hat{L}_z$ to the Hamiltonian, where $\hat{L}_z$ is the operator associated to the third component of the angular momentum) and then employing the imaginary-time method \cite{Marti_Pi,Gallemi_Stringari}. The starting condition for the imaginary-time dynamics is such that species-\textit{a} hosts a vortex pair while species-\textit{b} is localized (two narrow Gaussian distributions) at the vortex cores. As the fictitious dynamics advances, the position of the vortex cores, their healing length, together with the spatial distribution of species-\textit{b} atoms is iteratively self-consistently refined, until convergence is reached. We conclude this Section by noticing that, for our purposes, solutions of Eqs. (\ref{eq:Eig_prob}) of the type vortex/bright-soliton pairs (see, e.g. Fig. \ref{fig:Stationary_state}) could be either actual ground states or excited metastable states (i.e. either global or local minima in the energy landscape).

\section{Massive vortex pairs in a binary mixture of BECs}
\label{sec:Vortices_BECs}
\begin{figure}[h]
\centering
\includegraphics[width=1\linewidth]{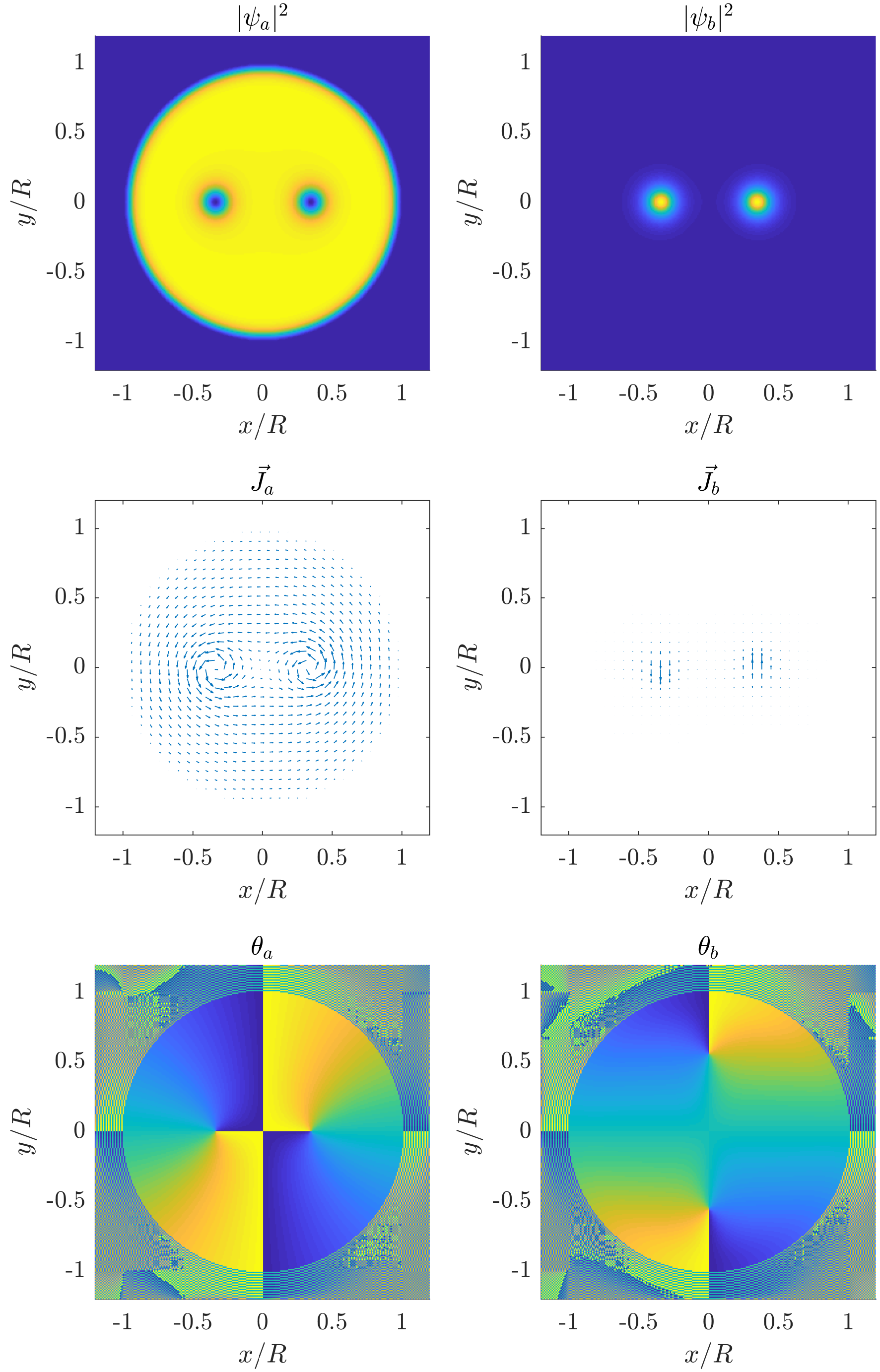}
\caption{Typical minimum-energy solution of eigensystem (\ref{eq:Eig_prob}). First (second) row corresponds to the square moduli [mass current density (see Eq. \ref{eq:Mass_current_density})] of the eigensolutions $\psi_a$ and $\psi_b$. In the first row, yellow (blue) is associated to large (zero) values of the density $|\psi|^2$. Third row corresponds to the \textit{phase}-field associated to $\psi_a$ and $\psi_b$ [blue (yellow) color corresponds to $-\pi$ ($+\pi$)]. Left (right) column corresponds to species $a$ ($b$). The following parameters have been used: $N_a=5\times 10^4$, $N_b = 10^3$, $\Omega =5$ rad/s, $R = $ 50 $\mu$m, $m_a=3.82 \times 10^{-26}$ kg, $m_b=6.48 \times 10^{-26}$ kg, $g_a= 52\times (4\pi\hbar^2 a_0)/m_a$, $g_b=7.6 \times (4\pi\hbar^2 a_0)/m_b$, $g_{ab}=24.2\times (2\pi\hbar^2 a_0)/m_{ab}$, $\ell_z=2$ $\mu$m. }
\label{fig:Stationary_state}
\end{figure}

Eigensystem (\ref{eq:Eig_prob}) was solved sweeping model parameter $N_b$, the number of species-\textit{b} atoms, which constitute the massive cores of species-\textit{a} vortices. As explained in Sec. \ref{sec:Numerical_model}, a suitable ansatz for the starting condition of the imaginary-time dynamics was chosen. In the whole range of $N_b$ that we explored (i.e. $N_b \,\in\, [5,\,1000]$), our numerical simulations \cite{HPC_Polito} converged to a stationary state of the type illustrated in Fig. \ref{fig:Stationary_state}.  
Basically, condensate \textit{a} is highly confined by the box-like potential (whose radius is $R=50\, \mu\text{m}$) and it is marked by the presence of two corotating vortices. The latter are symmetrically positioned with respect to center of the trap, they are such that the density $|\psi_a|^2$ goes to zero in the center of the cores and feature a quantized circulation. On the other hand, component \textit{b} occupies the vortex cores which, in turn, constitute an effective double-well potential for species-\textit{b} atoms \cite{Vortices_double_well}. 

One can gain a further insight into the discussed eigensolution by computing the mass current density 
\begin{equation}
    \label{eq:Mass_current_density}
    \vec{J}_c=  -\frac{i\hbar}{2}(\psi_c^*\nabla\psi_c-\psi_c\nabla\psi_c^*)
\end{equation}
(with $c=a,\,b$) which also corresponds to the momentum-per-particle distribution. As illustrated in the second row of Fig. \ref{fig:Stationary_state}, both vortices in species \textit{a} rotate anticlockwise, thus determining a collective motion of precession which is anticlockwise too. As concerns species-\textit{b} atoms, they are dragged by condensate \textit{a} and remain bound within the vortex cores thus featuring their same motion of precession around the center of the trap. With reference to the middle right panel of Fig. \ref{fig:Stationary_state}, one can appreciate that the left (right) peak of $|\psi_b|^2$ is translating downward (upward), i.e. along a direction tangential to the precession orbit.

Eventually, another important information that can be extracted from the eigensolutions $\psi_a$ and $\psi_b$ of equations (\ref{eq:Eig_prob}) concerns the \textit{phase} fields. The latter, denoted by $\theta_a$ and $\theta_b$, and such that $\psi_c=\sqrt{|\psi_c|^2}e^{i\theta_c}$ (where $c=a,\,b$), have been plotted in the lower panels of Fig. \ref{fig:Stationary_state}. As expected, $\theta_a$ features \textit{singularities} in correspondence of the vortices' centers, while, less expectedly, $\theta_b$ features singularities too. Nevertheless, the latter are found \textit{far} from the peaks of $|\psi_b|^2$. According to the basic properties of quantum fluids, the circulation 
$$
\mathcal{C}_\gamma[\vec{v}_c]=\oint_\gamma \vec{v}_c\cdot \mathrm{d}\vec{r}, \qquad (c=a,\,b) 
$$
of the velocity vector field associated to $\psi_c$, that means $\vec{v}_c=\frac{\hbar}{m_c}\nabla \theta_c $ (with $c=a,\,b$), is zero if the closed path $\gamma$ does \textit{not} encircle singularities of the phase field $\theta_c$ (with $c=a,\,b$). Conversely, due to the Feynman-Onsager quantization rules, it takes values $n\, h/m_c$, where $n\in \mathbb{Z}$, if one or more singularities of the associated field $\theta_c$ are encircled by $\gamma$. This information will be useful to better understand the discussion about the angular momentum of species-$b$ bosons (see Sec. \ref{sub:Angular_momentum_b}).

\subsection{Mass of the cores and equilibrium distance}
\label{sub:Equilibrium_distance}
\begin{figure}[h]
\centering
\includegraphics[width=0.8\linewidth]{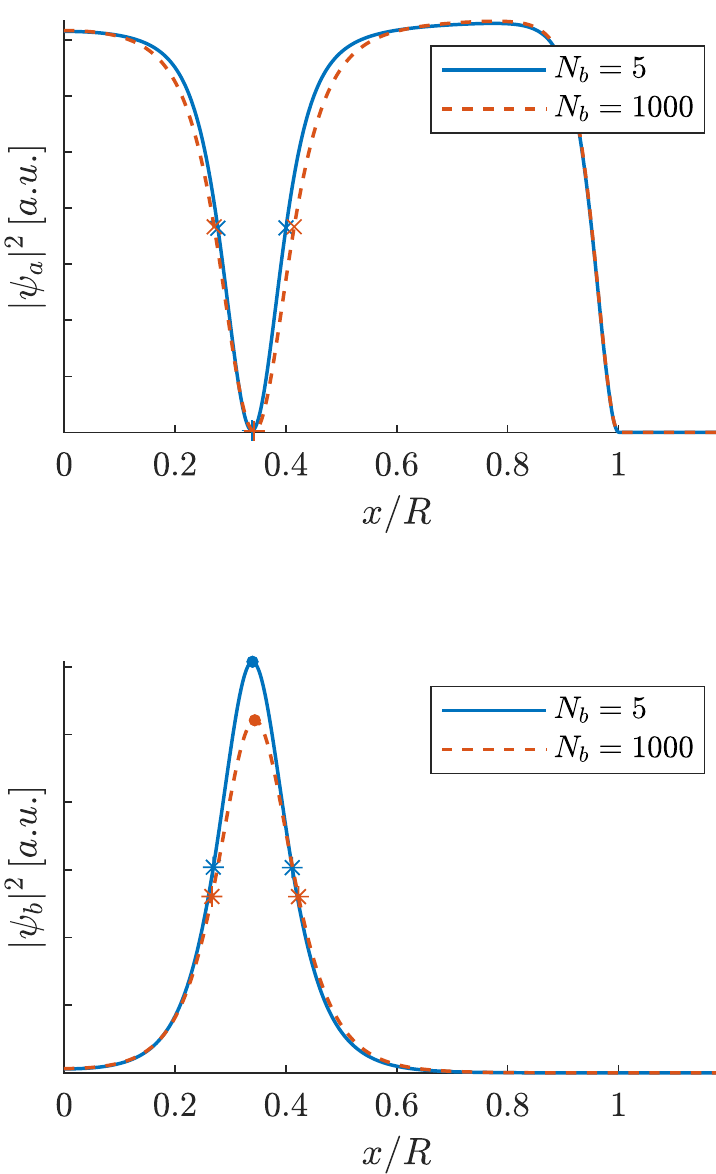}
\caption{Density profile of the minimum-energy solutions of eigensystem (\ref{eq:Eig_prob}) along the axis $y=0$ (we have plotted just the range $x>0$ because both $|\psi_a|^2$ and $|\psi_b|^2$ are symmetric with respect to $x=0$) for two different values of $N_b$. Upper panel: the position of markers `$+$' corresponds to $d_{vor}/2$, while the distance between markers `$\times$' corresponds to $2 \xi_a$. Lower panel: the position of markers `$\mathbf{\times}$' corresponds to $d_{peak}/2$, while the distance between markers `$*$' corresponds to $2 \sigma_b$. The following parameters have been used: $N_a=5\times 10^4$, $\Omega =5$ rad/s, $R = $ 50 $\mu$m, $m_a=3.82 \times 10^{-26}$ kg, $m_b=6.48\times 10^{-26}$ kg, $g_a= 52\times (4\pi\hbar^2 a_0)/m_a$, $g_b=7.6 \times (4\pi\hbar^2 a_0)/m_b$, $g_{ab}=24.2 \times (2\pi\hbar^2 a_0)/m_{ab}$, $\ell_z=2$ $\mu$m. }
\label{fig:Profile_comparison}
\end{figure}

Increasing the number of species-\textit{b} atoms (within the investigated range $N_b \,\in\, [5,\,1000]$), the distance $d_{vor}$ between the centers of the vortices \textit{increases}. Similarly, the distance between the two peaks of $|\psi_b|^2$, $d_{peak}$, increases upon increasing $N_b$. Fig. \ref{fig:Profile_comparison} shows how the presence of massive cores deforms and displaces the vortices. Notice that the position of the peaks of $|\psi_b|^2$ does not exactly match that of the minima of $|\psi_a|^2$ due to the centrifugal force on species-\textit{b} atoms and the finite repulsive coupling ($g_{ab}<+\infty$) between the two fluids which, in turn, allows for a non-zero penetration of fluid \textit{b} into fluid \textit{a}. Therefore, observables $d_{vor}$ and $d_{peak}$, which in the analytical model based on point-like vortices and point-like massive cores (see Sec. \ref{sec:Analytical_model}) collapse on the same variable ($d$), when estimated from the numerical solution of Eqs. (\ref{eq:Eig_prob}), despite being closely related, do not necessarily coincide.   

\begin{figure}[h]
\centering
\includegraphics[width=1\linewidth]{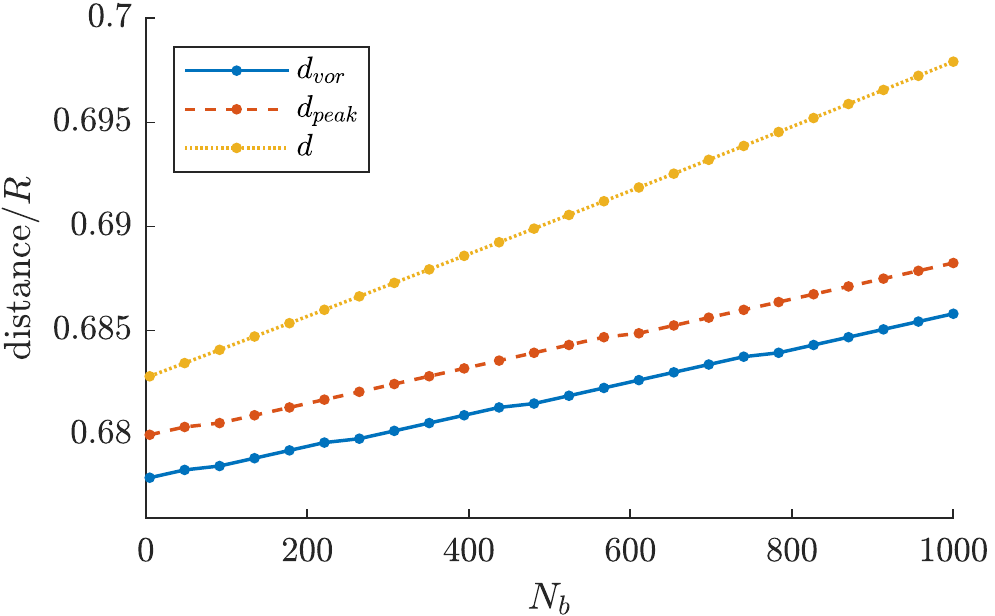}
\caption{Equilibrium distance: comparison between numerical ($d_{vor}$ and $d_{peak}$) and analytical ($d$) results. The following parameters have been used: $N_a=5\times 10^4$, $N_b\,\in\,[5,1000]$, $\Omega =5$ rad/s, $R = $ 50 $\mu$m, $m_a=3.82 \times 10^{-26}$ kg, $m_b=6.48 \times 10^{-26}$ kg, $g_a= 52\times (4\pi\hbar^2 a_0)/m_a$, $g_b=7.6 \times (4\pi\hbar^2 a_0)/m_b$, $g_{ab}=24.2\times (2\pi\hbar^2 a_0)/m_{ab}$, $\ell_z=2$ $\mu$m. }
\label{fig:Sweep_base}
\end{figure}

The functional dependence of $d_{vor}$ and $d_{peak}$ [extracted from the numerical solutions of system (\ref{eq:Eig_prob})] on $N_b$ is illustrated in Fig. \ref{fig:Sweep_base}, together with the relation $d(N_b)$, obtained, in turn, by means of substitutions
\begin{equation}
    \label{eq:Conversion_relations}
    k=\frac{h}{m_a}, \qquad \rho_*=\frac{N_a m_a}{\pi R^2 }, \qquad m=\frac{N_b m_b}{2}
\end{equation}
into Eq. (\ref{eq:Equi_distance_implicit}). Relations (\ref{eq:Conversion_relations}) allow one to match the analytical model (\ref{eq:Equi_distance_implicit}) based on point-like vortices and point-like massive cores with the actual parameters used to model the quantum fluids within the mean-field approach [see system (\ref{eq:Eig_prob})].  

The agreement between the analytical prevision (yellow dotted line) and numerical results (blue solid and red dashed lines) is remarkably good, both qualitatively (same quasi-linear functional dependence on $N_b$) and quantitatively (offset $<2\%$). Moreover, we would like to mention that numerical results (namely, the slope and the vertical shift of the corresponding lines of Fig. \ref{fig:Sweep_base}) can be shown to further approach the analytical prevision upon increasing $N_a$ and/or diminishing $g_b$, two changes that result in narrower cores and, therefore, in a scenario where Eq. (\ref{eq:Equi_distance_implicit}), based on our point-like approximation, reliably describes the mixture vortex state.

\subsection{Competition between centrifugal force and interspecies repulsion}
\label{sub:Effect_of_g_ab}
\begin{figure}[h]
\centering
\includegraphics[width=1\linewidth]{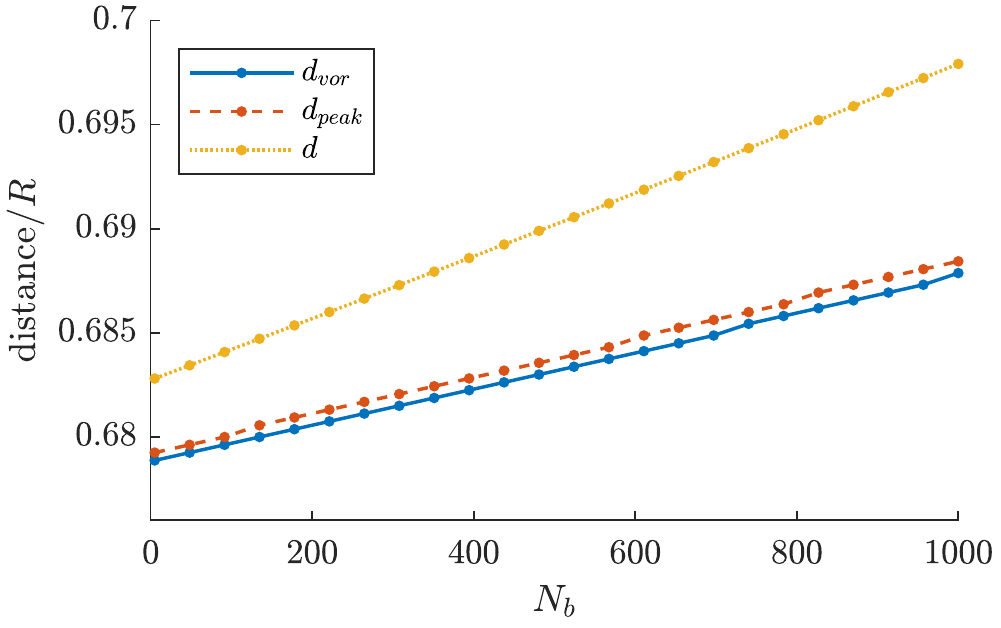}
\caption{Equilibrium distance: comparison between numerical ($d_{vor}$ and $d_{peak}$) and analytical ($d$) results. For these simulations, parameter $g_{ab}$ is $2.5$ times bigger than the one used  for Fig. \ref{fig:Sweep_base}, all the others being unchanged. }
\label{fig:Sweep_more_g_ab}
\end{figure}

As already mentioned, $d_{peak}$, although closely related to $d_{vor}$, is always slightly bigger than the latter. The motion of precession of the vortices around the center of the trap is responsible, in fact, for a centrifugal force on species-\textit{b} atoms which are, therefore, pushed outwards. This tendency is only partially opposed by the repulsive interaction between the two quantum fluids and it is the competition between these two forces what determines the exact values of $d_{vor}$ and $d_{peak}$. Increasing the interspecies repulsion $g_{ab}$, fluid \textit{a} gets more impenetrable to species-\textit{b} atoms, which therefore prove to be more tightly bound within the valleys of $|\psi_a|^2$. As a result of this increased reaction to the centrifugal force, the difference $d_{peak}-d_{vor}$ is remarkably smaller, as illustrated in Fig. \ref{fig:Sweep_more_g_ab} (where $g_{ab}$ has been set $2.5$ times bigger than the value used for Fig. \ref{fig:Sweep_base}).

\subsection{Stability beyond the immiscibility condition} 
We remark that the results illustrated in this article are obtained under the assumption that the two quantum fluids are \textit{immiscible}, meaning that their intra- and interspecies coupling parameters are such that $g_{ab}/\sqrt{g_a g_b}>1$. Actually, we should mention that vortex/bright-soliton complexes prove to be rather robust composite objects, as their stability extends also \textit{beyond} the immiscible regime. We have verified this by numerically solving Eqs. (\ref{eq:Eig_prob}), thus extending the results of Ref. \cite{Hamner}, derived in the context of dark-bright soliton complexes, to the case of vortex/bright-soliton complexes. The obtained density distributions, $|\psi_a|^2$ and $|\psi_b|^2$, are qualitatively similar to the ones illustrated in Fig. 1 and obtained in the case of \textit{immiscible} components. Of course, the difference between the miscible and the immiscible regime is that, in the latter case, bright solitons are more tightly confined within the vortex cores than in the former case. Moreover, due to the outward shift of the bright soliton from the vortex center, ensuing from the centrifugal force, the soliton itself, at a given time, feels an effective \textit{elliptical} potential which, in turn, deforms its circular shape. Concerning our sample, which is \textit{immiscible}, the deviation of the shape of bright solitons from a perfectly circular one can be neglected at first approximation, as the associated ellipticity, although being a decreasing function of $N_b$, is always $\approx 0.98$ in the whole considered range of $N_b$. 

When the two quantum fluids are miscible, in fact, bright solitons manage to invade the majority component in a more significant way. As a consequence, they play the role of less rigid (i.e. softer) massive cores. If, on one hand, this circumstance extends the robustness of these composite objects to the case of \textit{miscible} quantum fluids, on the other hand, in the miscible regime, our point-like model (\ref{eq:Lagrangian}) partially looses its validity, as the mass of the bright solitons is not concentrated in the vortices centers any more, but it spreads and occupies more extended spatial regions.

\section{Angular momentum of vortices and cores}
\label{sec:Angular_momentum}
This section is devoted to the analysis of the angular momentum of each component, an investigation that can offer a deeper insight into the Physics of the system. In particular, we show that the two massive cores (made of species-\textit{b} atoms) \textit{orbit} around the center of the trap, being dragged by the motion of precession of the vortices. Nevertheless, they do not \textit{rotate}, i.e. their orientation remains constant while they revolve.   
\subsection{Angular momentum of condensate \textit{a} }
\label{sub:Angular_momentum_a}
The angular momentum (per particle, in units of $\hbar$) of condensate \textit{a} can be computed as 
\begin{equation}
    \label{eq:L_z_a}
     \frac{\langle L_{z,a}\rangle }{N_a \hbar}=  -i \int \psi_a^* \left(x\frac{\partial}{\partial y} - y\frac{\partial}{\partial x} \right) \psi_a  \mathrm{d} x \, \mathrm{d}y 
\end{equation}
This quantity can be evaluated numerically from the solution of Eqs. (\ref{eq:Eig_prob}). 

On the other hand, it can also be estimated by means of a fully-analytical approach. Along the same lines discussed in Ref. \cite{Muntsa_off_axis_vortices} (where the authors investigated the case of \textit{harmonic} confinement), in fact, it is possible to derive the following expression:  
\begin{equation}
    \label{eq:L_z_a_tilde}
   \frac{\tilde{L}_{z,a}}{N_a \hbar}=  2\int_{r_{vor}}^{R}\,  2\pi  r \frac{1}{\pi R^2} \,\mathrm{d}r = 2\left[1- \left(\frac{r_{vor}}{R}\right)^2\right],
\end{equation}
where $r_{vor}:=d_{vor}/2$ constitutes the orbit radius.

\begin{figure}[h]
\centering
\includegraphics[width=1\linewidth]{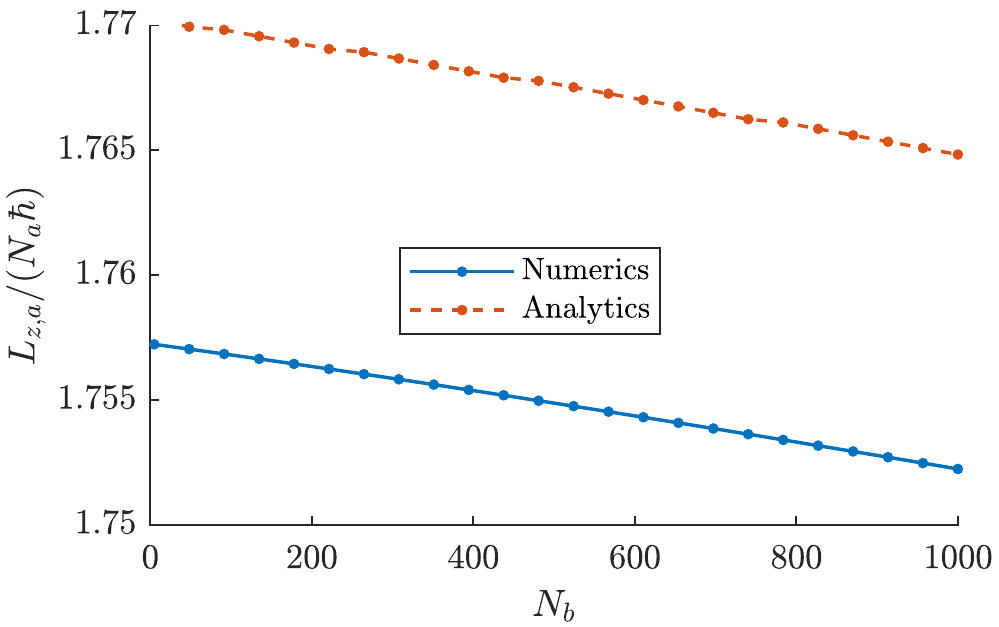}
\caption{Angular momentum (per particle, in units of $\hbar$) of condensate \textit{a}: comparison between numerical [see Eq. (\ref{eq:L_z_a})] and analytical [see Eq. (\ref{eq:L_z_a_tilde})] results. The following parameters have been used: $N_a=5\times 10^4$, $N_b\,\in\,[5,1000]$, $\Omega =5$ rad/s, $R = $ 50 $\mu$m, $m_a=3.82 \times 10^{-26}$ kg, $m_b=6.48 \times 10^{-26}$ kg, $g_a= 52\times (4\pi\hbar^2 a_0)/m_a$, $g_b=7.6 \times (4\pi\hbar^2 a_0)/m_b$, $g_{ab}=24.2\times (2\pi\hbar^2 a_0)/m_{ab}$, $\ell_z=2$ $\mu$m. }
\label{fig:L_z_a}
\end{figure}

As shown in Fig. \ref{fig:L_z_a}, Eq. (\ref{eq:L_z_a_tilde}) well fits the numerical data obtained by means of Eq. (\ref{eq:L_z_a}), the mismatch being $<0.8\%$. In this regard, it can be shown that the fitting accuracy further increases if one increases $N_a$ and/or decreases $g_b$, because, in this case, the point-like approximation of vortices and cores gets increasingly valid.

\subsection{Angular momentum of condensate \textit{b} }
\label{sub:Angular_momentum_b}
The angular momentum (per particle, in units of $\hbar$) of component \textit{b} can be analogously computed as 
\begin{equation}
    \label{eq:L_z_b}
     \frac{\langle L_{z,b}\rangle }{N_b \hbar}=  -i \int \psi_b^* \left(x\frac{\partial}{\partial y} - y\frac{\partial}{\partial x} \right) \psi_b  \mathrm{d} x \, \mathrm{d}y, 
\end{equation}
a quantity that can be evaluated numerically on the basis of the solution of Eqs. (\ref{eq:Eig_prob}).

As already mentioned, the two species-\textit{b} cores orbit around the center of the trap but they do not rotate around their own centers of mass. To prove this statement, we proceed along three different lines. 

\begin{figure}[h]
\centering
\includegraphics[width=0.8\linewidth]{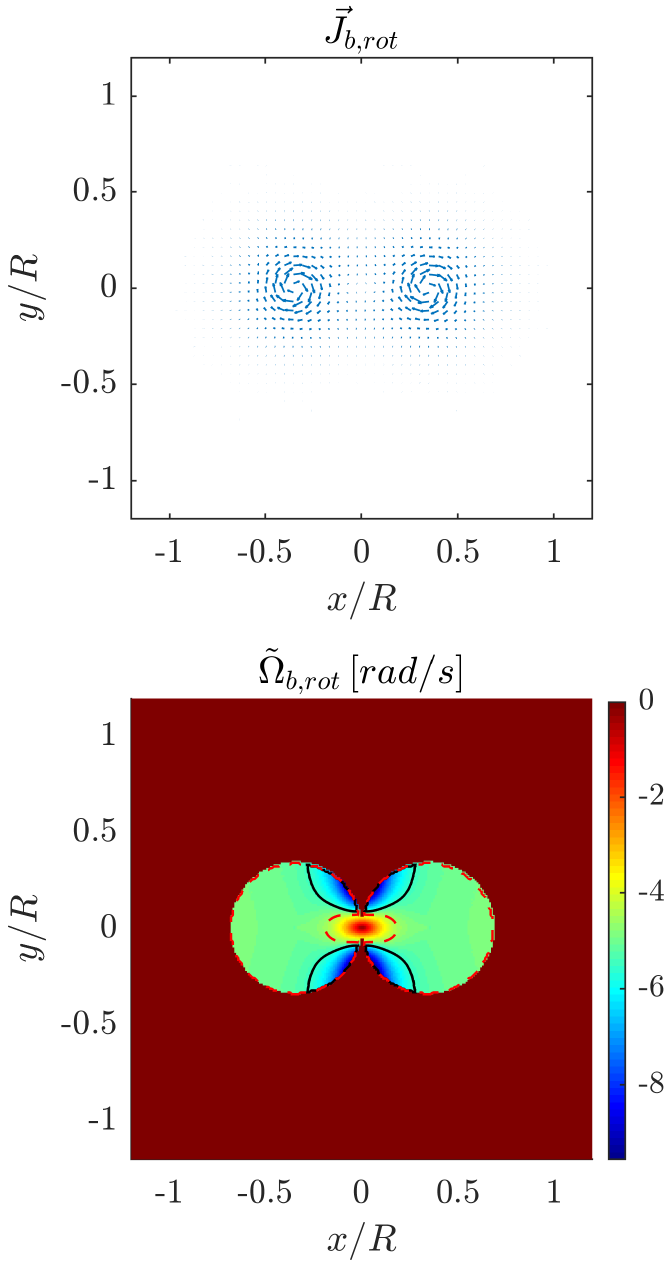}
\caption{Upper panel: species-\textit{b} mass current density in the rotating frame [see Eq. (\ref{eq:J_b_rot}) and the relevant explanation]: one can appreciate the two cores rotate \textit{clockwise}. Lower panel: species-\textit{b} local angular velocity, as defined by Eq. (\ref{eq:Local_Omega}); The solid black and red dashed lines correspond to $\tilde{\Omega}_{b,rot}=-5\pm0.5$ rad/s and have been drawn to illustrate that the two species-\textit{b} cores indeed rotate as two (almost) rigid bodies (see discussion in the main text). The following parameters have been used: $N_a=5\times 10^4$, $N_b = 10^3$, $\Omega =5$ rad/s, $R = $ 50 $\mu$m, $m_a=3.82 \times 10^{-26}$ kg, $m_b=6.48 \times 10^{-26}$ kg, $g_a= 52\times (4\pi\hbar^2 a_0)/m_a$, $g_b=7.6 \times (4\pi\hbar^2 a_0)/m_b$, $g_{ab}=24.2\times (2\pi\hbar^2 a_0)/m_{ab}$, $\ell_z=2$ $\mu$m. }
\label{fig:Rigid_body}
\end{figure}

\paragraph{Mass current density in the rotating frame.} In the lab frame, it is possible to compute the mass current density $\vec{J}_b$ associated to $\psi_b$ [see Eq. (\ref{eq:Mass_current_density})]. The corresponding vector field is illustrated in the middle right panel of Fig. \ref{fig:Stationary_state}. It is clear that the left (right) core is moving downward (upward), dragged by the anticlockwise motion of precession of the vortices. Due to the characteristic magnitude of $|\vec{J}_b|$, this plot does not allow one to understand whether the cores change their orientation or not along their orbit around the center of the trap. To circumvent this limitation, we have computed the species-\textit{b} mass current density in a (non-inertial) frame rotating with the same angular velocity $\Omega$ distinguishing the motion of precession of the vortices. More specifically, in the rotating frame, $\vec{J}_{b,rot}$ reads
\begin{equation}
    \label{eq:J_b_rot}
    \vec{J}_{b,rot}=m_b |\psi_b|^2\vec{v}_{b,rot},
\end{equation}
where $\vec{v}_{b,rot}=\vec{v}_b-\vec{V}$ with $\vec{v}_b =\frac{\vec{J}_b}{m_b|\psi_b|^2}$ and $\vec{V}=\vec{\Omega}\wedge \vec{r}$. $\vec{J}_b$ is numerically computed through Eq. (\ref{eq:Mass_current_density}). Notice also that, at $t=0$, $|\psi_b|^2 \equiv |\psi_{b,rot}|^2$, thus justifying its use in Eq. (\ref{eq:J_b_rot}). The result of this procedure is illustrated in the upper panel of Fig. \ref{fig:Rigid_body} which shows that the two species-\textit{b} cores, when observed from the rotating frame, rotate around their respective centers of mass, with angular velocity $-\Omega$ (the minus sign being due to the \textit{clockwise} direction).
On top of that, we have evidenced how these two cores rotate almost as if they were rigid bodies, meaning that the (absolute value of the) velocity field $\vec{v}_{b,rot}$ around each center of mass linearly increases with the distance $r_C$ from the respective center of mass (this is not in contrast with the irrotational property of quantum gases, as the new reference frame is not inertial). In view of the symmetry of the ground-state configuration depicted in Fig. \ref{fig:Stationary_state}, we refer to the left (right) center of mass when considering the velocity field in the half-plane $x<0$ ($x>0$). The lower panel of Fig. \ref{fig:Rigid_body} shows the local angular velocity of species-\textit{b} cores when observed from the rotating frame. This quantity, defined as 
\begin{equation}
    \label{eq:Local_Omega}
    \tilde\Omega_{b,rot}(x,y) =-\frac{|\vec{v}_{b,rot}|}{r_C},
\end{equation}
[where $r_C$ is the distance of point $(x,y)$ from the left (right) center of mass when $x<0$ ($x>0$)], takes the (almost) constant value $\approx -5$ rad/s in the most part of the regions where $|\psi_b|^2$ is non-zero.

As an alternative indicator, in order to investigate the possible rotational properties of bright solitons, one could employ the vorticity distribution in the rotating frame $\vec{w}_{b,rot}=\vec{\nabla}\wedge \vec{v}_{b,rot}$. Apart from two Dirac-delta-like singularities exactly where phase singularities are (see lower right panel of Fig. \ref{fig:Stationary_state}), one would observe two quasi-plateaus at $|\vec{w}_{b,rot}|=2\tilde{\Omega}_{b,rot}\approx -10\, \mathrm{rad/s}$. This circumstance is in great agreement with the fact that the vorticity distribution associated to a rigid body rotating with angular frequency $\Omega$ is uniform and  equal to $2\Omega$.

In conclusion, we have proved that, in the (non-inertial) rotating frame, the two species-\textit{b} cores rotate around their respective centers of mass with (almost uniform) angular velocity $-\Omega$. This allows us to conclude that the they keep their orientation fixed when observed from the lab frame.

\begin{figure}[h]
\centering
\includegraphics[width=1\linewidth]{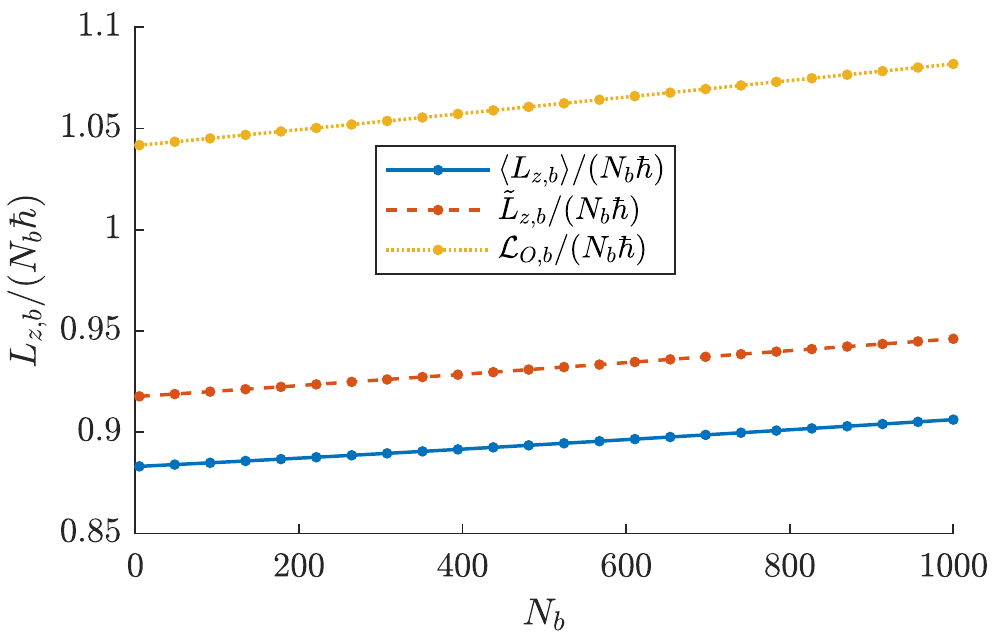}
\caption{Angular momentum (per particle, in units of $\hbar$) of component \textit{b}: comparison between numerical [solid blue line, associated to Eq. (\ref{eq:L_z_b})] and semi-analytical [red dashed line, corresponding to Eq. (\ref{eq:L_z_b_tilde})] results [the yellow dotted line, associated to Eq. (\ref{eq:Contribution_1}) represents the angular momentum of the system if its motion was a pure revolution around the center of the trap $O$]. The following parameters have been used: $N_a=5\times 10^4$, $N_b\,\in\,[5,1000]$, $\Omega =5$ rad/s, $R = $ 50 $\mu$m, $m_a=3.82 \times 10^{-26}$ kg, $m_b=6.48 \times 10^{-26}$ kg, $g_a= 52\times (4\pi\hbar^2 a_0)/m_a$, $g_b=7.6 \times (4\pi\hbar^2 a_0)/m_b$, $g_{ab}=24.2\times (2\pi\hbar^2 a_0)/m_{ab}$, $\ell_z=2$ $\mu$m. }
\label{fig:L_z_b}
\end{figure}

\paragraph{Analytical estimate of the angular momentum.} To corroborate what elucidated in the previous paragraph, we show that the functional dependence of quantity (\ref{eq:L_z_b}) on model parameter $N_b$ can be well fitted by the semi-analytical model
\begin{equation}
    \label{eq:L_z_b_tilde}
    \frac{\tilde{L}_{z,b}}{N_b \hbar}\approx \frac{\tilde{\mathcal{L}}_{O,b}}{N_b \hbar} + \frac{\tilde{\mathcal{S}}_{C_L,b}}{N_b \hbar}+
   \frac{\tilde{\mathcal{S}}_{C_R,b}}{N_b \hbar},
\end{equation}
where
\begin{equation}
\label{eq:Contribution_1}
   \frac{\tilde{\mathcal{L}}_{O,b}}{N_b \hbar} = \frac{\Omega}{\hbar} \int\, m_b |\psi_{b}|^2 r^2\, \mathrm{d} x \, \mathrm{d}y,
\end{equation}
and where terms
\begin{equation}
    \label{eq:Contribution_2}
    \frac{\tilde{\mathcal{S}}_{C_L,b}}{N_b \hbar} = -\frac{\Omega}{\hbar} \int\, m_b |\psi_{b}|^2\Theta(-x) r_{C_L}^2\, \mathrm{d} x \, \mathrm{d}y ,
\end{equation}
\begin{equation}
\label{eq:Contribution_3}
    \frac{\tilde{\mathcal{S}}_{C_R,b}}{N_b \hbar} = -\frac{\Omega}{\hbar} \int\, m_b |\psi_{b}|^2\Theta(x) r_{C_R}^2\, \mathrm{d} x \, \mathrm{d}y 
\end{equation}
are introduced to take into account that, in the lab frame, the two species-\textit{b} cores revolve but keep their orientation fixed [Heaviside functions $\Theta(-x)$ and $\Theta(x)$ allow one to select the left and the right core respectively]. The integrals in expressions (\ref{eq:Contribution_1})-(\ref{eq:Contribution_3}) represent three moment of inertia corresponding, respectively, to the anticlockwise revolution of the whole system around the center of the trap $O$, and to the effective clockwise rotation of the left (right) core around its own center of mass $C_L$ ($C_R$) [in this regard, $r^2:=x^2+y^2$, $r_{C_\alpha}^2:=(x-x_{C_\alpha})^2+(y-y_{C_\alpha})^2$, with $\alpha=L,\,R$]. The latter effective motions indeed compensate for the fact that a \textit{pure} motion of revolution [captured by Eq. (\ref{eq:Contribution_1})] would determine a \textit{change} in the orientation of the cores along the circular orbit. Figure \ref{fig:L_z_b} shows a very good agreement between Eq. (\ref{eq:L_z_b}) and formula (\ref{eq:L_z_b_tilde}), the error being always $<4\%$.

\paragraph{Observation of the associated phase field.} As illustrated in the lower right panel of Fig. \ref{fig:Stationary_state}, the phase field associated to $\psi_b$, $\theta_b$, features two singularities \textit{far} from the centers of bright solitons. If one considers a closed path $\gamma$ encircling one of the bright solitons, since no singularities of $\theta_b$ are surrounded by it, the circulation $\mathcal{C}_\gamma[\vec{v}_b]$ of the velocity field $\vec{v}_b$ along $\gamma$ must be zero (see Sec. \ref{sec:Numerical_model}). This implies that both bright solitons do \textit{not} rotate around their own axes (in the laboratory inertial frame). On the other hand, these singularities in $\theta_b$ indeed play an important role because the corresponding velocity field $\vec{v}_b=\frac{\hbar}{m_b}\nabla \theta_b$ is what determines the collective precession motion of species-b bosons.

In summary, we notice that the interspecies repulsive coupling $g_{ab}$ is the interaction underlining the dragging of species-\textit{b} cores by species-\textit{a} vortices, which therefore exhibit the same motion of precession. In spite of this precession, species-$b$ cores keep their orientation constant (in the inertial frame of the laboratory). Due to the irrotational properties of quantum fluids, in fact, a soliton-like distribution cannot be put in rotation around its own axis (as if it was a rigid body) without creating a phase singularity at its center. On the other hand, the creation of such a phase singularity would turn the soliton-like original distribution into a vortical object. To conclude this Section, we would like to mention the possible existence of the Andreev-Bashkin effect \cite{Andreev0,Andreev1,Andreev2,Andreev3}, according to which, the mass current density $\vec{J}_i$ (with $i=a,\, b$) in one species will depend, in general, also on the velocity $\vec{v}_j$ (with $j=b,\,a$) of the other species. In other words, the condensate density $\rho_{ij}$ is a non-diagonal matrix, a circumstance which implies the relations
$$
   \vec{J}_a =\rho_{aa}\vec{v}_a + \rho_{ab}\vec{v}_b
$$
$$
   \vec{J}_b =\rho_{ba}\vec{v}_a + \rho_{bb}\vec{v}_b
$$
At the microscopic scale, this drag between mass current densities comes from the formation of quasi-particles with non-zero content of mass for either of the two components \cite{Andreev3}. As a consequence, the transport properties of the two quantum fluids turn out to be coupled: the flow of one species influences the mass transport in the other species \cite{Andreev0}. This effect is known to be rather elusive \cite{Andreev3} and, for our sample condition, we expect it to be negligible, as the off-diagonal matrix elements $\rho_{ab}$ and $\rho_{ba}$, which depend on the the overlap between $\psi_a$ and $\psi_b$, should be small, given that the two discussed fluids are \textit{immiscible}. In view of its complexity, the possible presence of this effect in the discussed system of orbiting vortex/bright-soliton complexes will be analyzed in a future work.

\section{Vortex healing lengths and size of the massive cores}
\label{sec:Healing_diameter}
The presence of species-\textit{b} massive cores within species-\textit{a} vortices affects the healing length of the latter. The intraspecies repulsive interaction, in fact, tends to enlarge the cores which, in turn, tend to swell (from the inside) the profile of the vortices because of the interspecies repulsive coupling. Flipping the perspective, the expansion of the cores is dammed by the species-\textit{a} fluid, which plays the role of an effective confining potential for species-\textit{b} atoms. 

In the attempt to estimate the equilibrium healing length $\xi_a$ of species-\textit{a} vortices and the equilibrium characteristic size of species-\textit{b} cores, $\sigma_b$, we present the following heuristic equations
$$
    \frac{\hbar^2}{2m_a}\frac{1}{\xi_a^2} = +g_a n_a - g_{ab} n_a n_b \pi (\sigma_b^2-\xi_a^2) \ell_z,
$$
\begin{equation}
    \label{eq:Heuristic_model}
    \frac{\hbar^2}{2m_b}\frac{1}{\sigma_b^2} =- g_b n_b + g_{ab} n_a n_b  \pi (\sigma_b^2-\xi_a^2) \ell_z,
\end{equation}
where 
$$
   n_a = \frac{N_a}{\pi r_{box}^2 \ell_z}, \qquad    n_b = \frac{N_b/2}{\pi \sigma_b^2 \ell_z}.
$$
Notice that the the first equation of system (\ref{eq:Heuristic_model}) reduces, in the case of no interspecies interaction, to
$$
   \xi_{a,0}=\sqrt{\frac{\hbar^2}{2m_a g_a n_a}},
$$
the well-known formula derived in the context of single-species vortices \cite{Stringari}. Similarly, the second equation, if $g_{ab}=0$, gets structurally similar to formula
$$
   \sigma_{b,0}=\sqrt{\frac{\hbar^2}{2m_b |g_b| n_{b,0}}},
$$
giving the characteristic size of a soliton in the case of attractive interactions \cite{Stringari} (of course, in this context, $n_{b,0}$ represents the central density). The extra term $g_{ab} n_a n_b  \pi (\sigma_b^2-\xi_a^2) \ell_z$ is introduced to take into account the interspecies repulsion, an interaction that manifests only in those regions where $\psi_a$ and $\psi_b$ overlap, i.e. only in the two annuli centered in the vortices' centers and whose outer and inner radii are $\sigma_b$ and $\xi_a$ respectively.    

In order to compare the previsions provided by equations (\ref{eq:Heuristic_model}) with the values extracted from the numerical solutions of system (\ref{eq:Eig_prob}), one has to give the operational definition of ``vortex healing length" and ``core characteristic size". From the numerical side, with reference to Fig. \ref{fig:Profile_comparison}, we agree to measure the half width at half maximum of the valley of $|\psi_a|^2$, $\lambda_a$, and the half width at half maximum of the peak of $|\psi_b|^2$, $\lambda_b$. From the analytical side, the estimates of quantities $\lambda_a$ and $\lambda_b$ are given by the solutions of system (\ref{eq:Heuristic_model}), $\xi_a$ and $\sigma_b$, multiplied by two suitable constant conversion factors, $1.30$ and $1.15$ respectively, which are determined from their numerical counterpart in the case $N_b=1$, that means in a scenario where species-\textit{b} cores have a negligible impact on species-\textit{a} vortices.

\begin{figure}[h]
\centering
\includegraphics[width=1\linewidth]{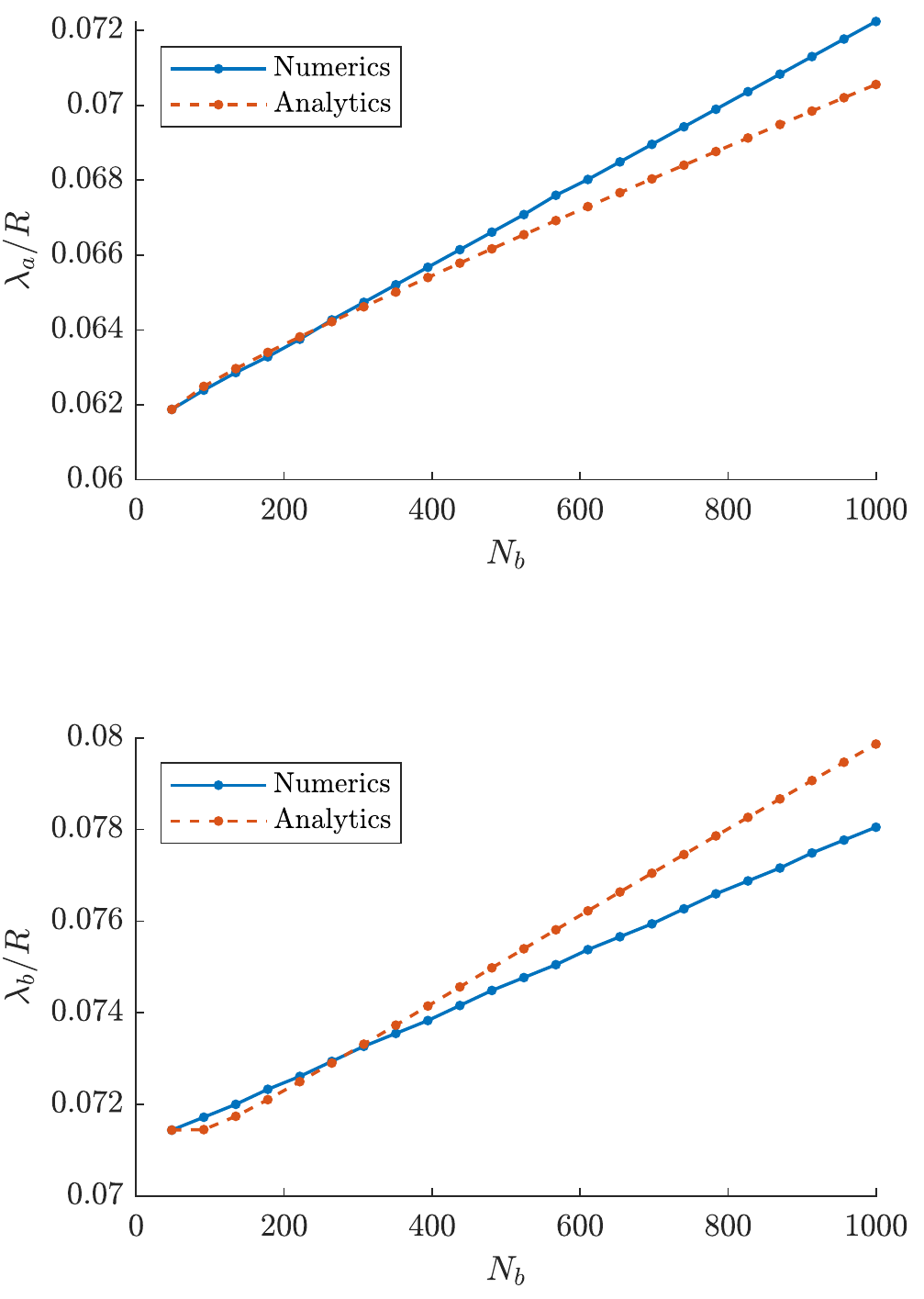}
\caption{Upper (lower) panel: comparison between the numerically- determined and the analytically- estimated half width half maximum, $\lambda_a$, of species-\textit{a} vortices (of species-\textit{b} cores, $\lambda_b$). The following parameters have been used: $N_a=5\times 10^4$, $N_b\,\in\,[1,1000]$, $\Omega =5$ rad/s, $R = $ 50 $\mu$m, $m_a=3.82 \times 10^{-26}$ kg, $m_b=6.48 \times 10^{-26}$ kg, $g_a= 52\times (4\pi\hbar^2 a_0)/m_a$, $g_b=7.6 \times (4\pi\hbar^2 a_0)/m_b$, $g_{ab}=24.2\times (2\pi\hbar^2 a_0)/m_{ab}$, $\ell_z=2$ $\mu$m. }
\label{fig:Healing_and_Sigma}
\end{figure}

As illustrated in Fig. \ref{fig:Healing_and_Sigma}, equations (\ref{eq:Heuristic_model}) well capture the functional dependence of $\lambda_a$ and of $\lambda_b$ on $N_b$. 

\section{Concluding remarks}
\label{sec:Conclusions}
In this work, we have investigated a notable class of configurations exhibited by a bosonic immiscible binary mixture loaded in a box-like circular trap, namely, minimum-energy states where species-\textit{b} atoms are trapped within the vortex cores of species-\textit{a} fluid. Both within a fully-analytical framework and by means of a systematic analysis of the numerical solutions of the associated two coupled GPEs, we have shown that the presence of massive cores alters the equilibrium distance distinguishing the motion of precession of the vortex pair. Interestingly, for the considered choices of model parameters (repulsive intra- and interspecies interactions such that, in the homogeneous case, the miscibility condition $g_{ab}<\sqrt{g_a g_b}$ is not met) the dynamical mean-field picture of the mixture has been shown to reduce to much simpler effective equations exhibiting an evident Lorentz-like magnetic form, where massive vortices play the role of massive charges confined on a plane and subject to a magnetic field. Species-\textit{b} cores, in turn, are dragged by fluid \textit{a} and thus follow their same motion of precession around the trap center; nevertheless, while orbiting, they keep their orientation constant, meaning that there is no tangential entrainment between the two fluids. We have also derived, in the context of the Thomas-Fermi approximation, a simple formula to estimate the angular momentum of condensate $\textit{a}$ and we have shown, by means of a suitable change of reference frame, that species-\textit{b} cores effectively behave as two rigid bodies. Eventually, we have introduced a system of heuristic but effective equations to estimate the characteristic size of vortices and cores hosted therein.

\section*{Acknowledgements}
\label{sec:Acknowledgements}
The authors wish to thank M. Pi, A. Gallem\'{i} and Y. Bidasyuk for their valuable help in setting up the numerical simulations. V. P. is indebted with F. Minardi for useful discussions about
trapping techniques. M. G. and R. M. acknowledge financial support from the Spanish MINECO and Fondo Europeo de Desarrollo Regional (FEDER, EU) under Grant No. FIS2017-87801-P, and from Generalitat de Catalunya Grant No. 2017SGR533.

\end{document}